\def\solmas{{M$_\odot$}}

\def\mnras{{MNRAS}}
\def\apj{{ApJ}}

\def\simless{\mathbin{\lower 3pt\hbox
   {$\rlap{\raise 5pt\hbox{$\char'074$}}\mathchar"7218$}}}   % < or of order
\def\simgreat{\mathbin{\lower 3pt\hbox
   {$\rlap{\raise 5pt\hbox{$\char'076$}}\mathchar"7218$}}}   % > or of order

\documentclass{iaus}
\usepackage{graphicx}

\title[spiral triggering] %% give here short title %%
{Spiral arm triggering of star formation}

\author[Bonnell \& Dobbs]   %% give here short author list %%
{Ian A. Bonnell$^1$%
  \and Clare L. Dobbs$^2$}

\affiliation{$^1$SUPA, School of Physics and Astronomy, University of St Andrews, KY16 9SS, UK \break email: iab1@st-and.ac.uk\\[\affilskip]
$^2$Department of Physics, University of Exeter, UK }

\pubyear{2006}
\volume{237}  %% insert here IAU Symposium No.
\pagerange{119--126}
\date{?? and in revised form ??}
\jname{Triggered star formation in a turbulent ISM}
\editors{B. Elmegreen \& J. Palous, eds.}
\begin{document}

\maketitle

\begin{abstract}

We present numerical simulations of the passage of clumpy gas through a galactic spiral shock, the 
subsequent formation of giant molecular clouds (GMCs) and the triggering of star formation. 
The spiral shock forms dense clouds while dissipating kinetic energy, producing 
regions that are locally gravitationally bound and collapse to form stars. In addition 
to triggering the star formation process, the clumpy gas passing through the 
shock naturally generates the observed velocity dispersion size relation of molecular clouds. 
In this scenario, the internal motions of GMCs need not be turbulent in nature.
The coupling 
of the clouds' internal kinematics to their externally triggered formation 
removes the need for the clouds to be self-gravitating. Globally unbound molecular clouds 
provides a simple explanation of 
the low efficiency of star formation. While dense regions in the shock become 
bound and collapse to form stars, the majority of the gas disperses as it leaves the spiral arm.

%\keywords{Keyword1, keyword2, keyword3, etc.}
\end{abstract}

\firstsection % if your document starts with a section,
\section{Introduction}

Star formation has long been known to occur primarily in the spiral arms of disc
 galaxies (Baade 1963). Spiral arms are denoted by the presence of young stars, HII
 regions, dust and giant molecular clouds, all signatures of the star formation
 process (Elmegreen \& Elmegreen~1983; Ferguson \etal 1998). 
What is still unclear is the exact role of the spiral arms in inducing the star formation. 
Is it simply that the higher surface density due to the orbit crossing is sufficient to initiate
star formation, as in a Schmidt law, or do the spiral arms play a more active role? 
Roberts (1969) first suggested that the spiral shock that occurs as the gas flows through the
potential minima  triggers the  star formation process in spiral galaxies. Shock dissipation
of excess kinetic energy can result in the formation of bound structures which then collapse
to form stars.

Giant molecular clouds (GMCs) are observed to contain highly supersonic
motions and a wealth of structure on all length scales (Larson~1981;
Blitz \& Williams 1999).  The supersonic
motions are generally thought to be 'turbulent' in nature and to be the cause
of the density structure in GMCs (Mac Low \& Klessen~2004; Elmegreen
\& Scalo~2004).  We propose an alternative scenario whereby it is the passage of the clumpy
interstellar medium through a galactic spiral shock that not only produces
the dense environment in which molecular clouds form (Cowie 1981; Elmegreen 1991),
but also gives
rise at the same time to their supersonic internal motions (Bonnell \etal 2006).

\section{Global Simulations}

\begin{figure*}[t]
\centering
\includegraphics[scale=0.5]{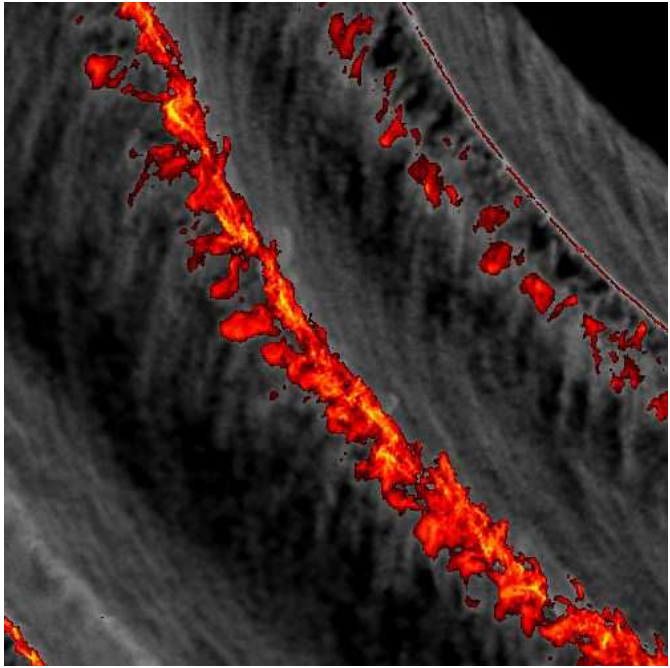}
\caption{The formation of molecular clouds is shown as the gas passes  through a spiral shock
(Dobbs \etal\ 2006).Note the spurs and feathering that appear as the dense clumps
are sheared away upon leaving the spiral  arm.}
\end{figure*}

Recent global simulations of non-self gravitating gas dynamics in spiral galaxies 
(Dobbs, Bonnell \& Pringle 2006)
show that the spiral shocks can account for the formation of molecular gas from cold ($T\simless100$ K)
atomic gas
and generate the large scale distribution of molecular clouds in spiral arms. Structures
in the spiral arms arise due to the shocks that tend to gather material together on 
converging orbits.
Thus, structures grow in time through multiple spiral arm passages. These structures present in
the spiral arms are also found to form the spurs and feathering in the interarm region as they are 
sheared by the divergent orbits when leaving the spiral arms (Dobbs \& Bonnell 2006a). 
The high gas densities that result
from the high Mach number shocks are sufficient for rapid formation of $H_2$ gas and thus
of giant molecular clouds. If, in contrast, the gas is warm ($T\ge1000$ K) when it enters the shock, 
then $H_2$ formation cannot occur due
to the lower gas densities in the shock. In this model, molecular clouds are limited
to spiral arms as it is only there that the gas is sufficiently dense to form molecules. These
clouds need not be self-gravitating as their formation is independent of self-gravity.
The velocity  dispersion in the gas also undergoes periodic bursts during the spiral arm passage as
the clumpy shock drives supersonic random motions into the gas (see below). Such bursts in the internal
gas motions are likely to be observable and would give support for a spiral shock origin of giant
molecular clouds and the triggering of star formation.

\begin{figure*}[t]
\centering
\includegraphics[scale=0.5]{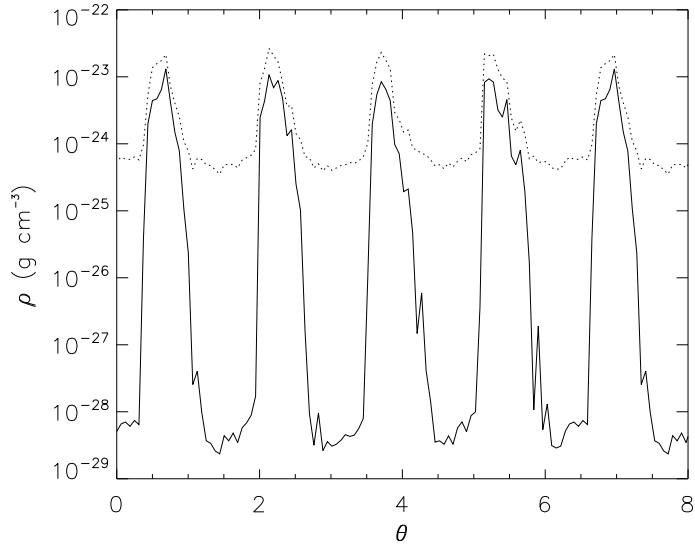}
\caption{The total gas (dotted) and $H_2$ (solid)  gas densities are plotted
agains azimuth for an annuli at 5 kpc. the gas is averaged over cells of 50 pc in size
(from Dobbs \etal 2006). Molecular gas is almost exclusively contained  .
}
\end{figure*}

\section{Triggering of Star Formation in the Spiral Shock}

In simulations where self-gravity is included, the passage of gas through a
spiral shock can result in the triggering of star formation (Bonnell \etal\ 2006).
The evolution, over 34 million years,  of $10^6$ \solmas\ of  gas  passing through
the spiral potential is shown in Figure~\ref{spiraltrig}. The initially
clumpy, low density gas ($\rho \approx 0.01$ \solmas pc$^{-3}$) 
is compressed by the spiral shock as it leaves
the minimum of the potential.  The shock forms some very dense ($>10^3$ \solmas pc$^{-3}$)
regions, which  become
gravitationally bound and thus collapse to form regions of star
formation. Further accretion onto these regions, modeled with sink-particles
in SPH (Bate \etal\ 1995),
raises their masses to that of typical stellar clusters ($10^2$ to
$10^4$ \solmas).  Star formation occurs within $2 \times 10^6$ years
after molecular cloud densities are reached. The total
spiral arm passage lasts for $\approx 2 \times 10^7$ years.  The gas
remains globally unbound throughout the simulation and re-expands in
the post-shock region. The star formation efficiency is of order 10 \%
and should be taken as an upper limit in the absence of any form of stellar feedback.

\begin{figure*}[t]
\centering
\includegraphics[scale=1.50]{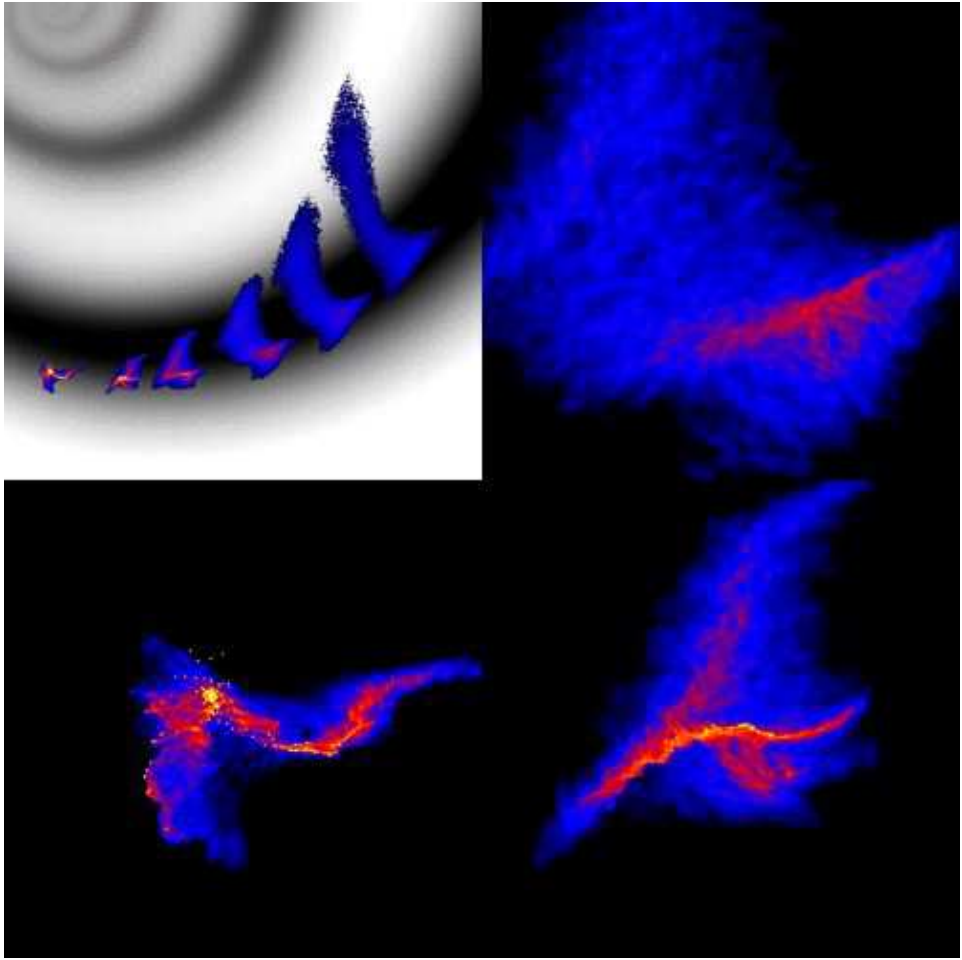}
\caption{The evolution of cold interstellar gas through a spiral arm is shown relative to the spiral potential of the
galaxy (upper left-panel). The minimum of the spiral
potential is shown as black and the overall galactic potential is not
shown for clarity. The 3 additional panels, arranged
clockwise, show close-ups of the gas as it  is compressed in the shock and subregions become self-gravitating.  Gravitational 
collapse and star formation occurs  within $2 \times 10^6$ years of the gas
reaching molecular cloud densities.  The
cloud produces stars inefficiently as the gas is not globally bound. 
}
\end{figure*}

The star forming clouds that form in the spiral shocks are generally unbound and thus
disperse once they leave the spiral arm. Numerical simulations of unbound clouds
(Clark \& Bonnell 2004; Clark \etal 2005)
have recently shown that they can form local subregions which are gravitationally unstable
and thus form stars.   In contrast, the bulk of the cloud does not become bound and thus disperses
without entering into the star formation process. The resultant star formation efficiencies
are of order 10 per cent, even
without the presence of the divergent flows of gas leaving a spiral arm. 
In fact, arbitrarily low star formation efficiencies are
possible with relatively small deviations from bound conditions.This then offers
a simple physical mechanism to explain the low star formation efficiency in our Galaxy.

\begin{figure*}[t]
\centering
\includegraphics[scale=0.5]{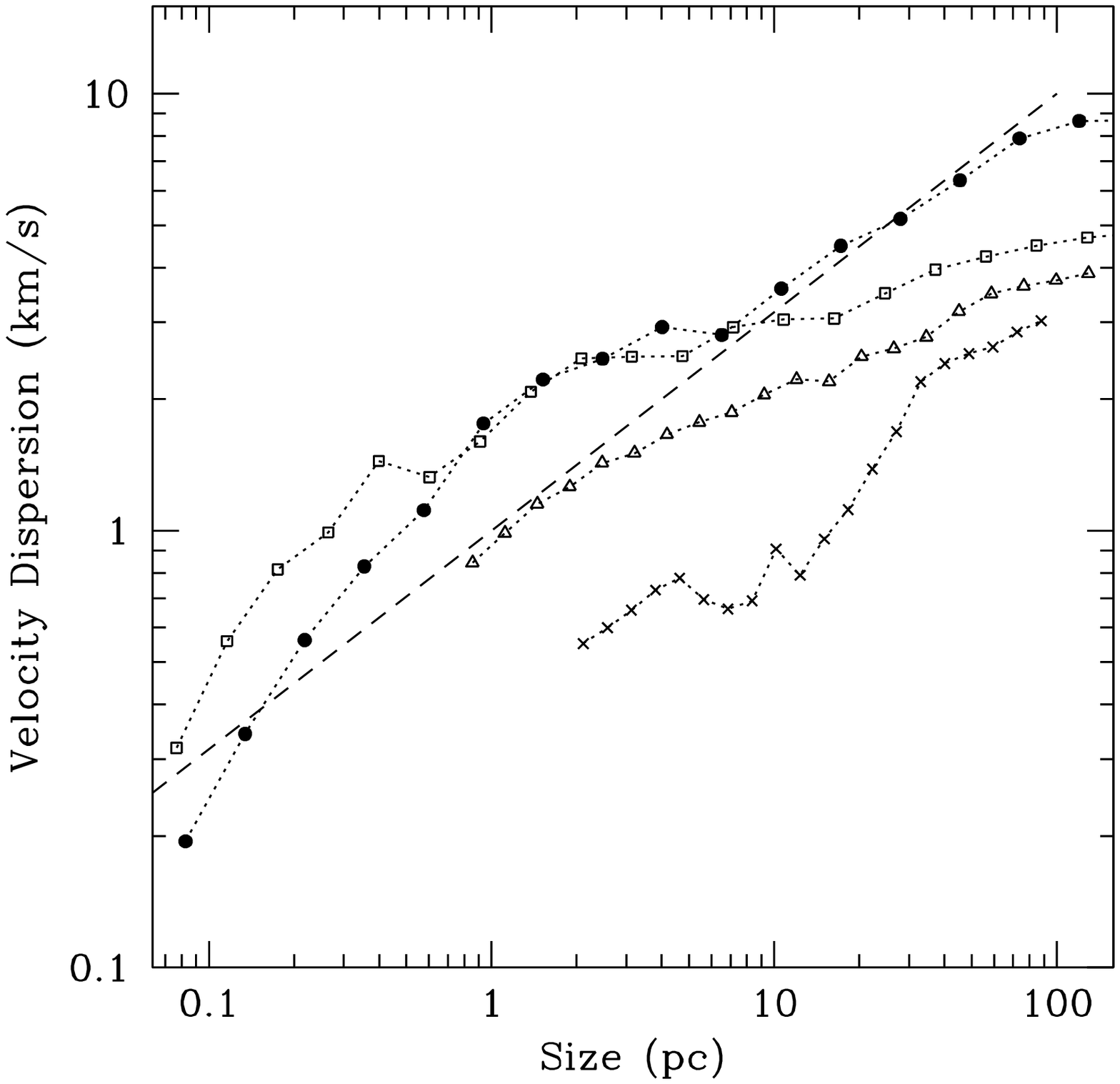}
\caption{The velocity dispersion is plotted as a
function of size at 5 different times during the passage of the gas
through a spiral shock. The velocity dispersion is
plotted at $4.2$, $14$, $18$, $23$, and $27 \times 10^6$ years  after the start of the simulation. Star formation is
initiated at $\approx 23 \times 10^6$ years. The dashed
line indicates the Larson relation for molecular clouds where $\sigma
\propto R^{-1/2}$ (Larson~1981; Heyer \& Brunt 2004).
}
\end{figure*}

\subsection{The Generation of the Internal Velocity Dispersion}

In addition to
triggering star formation, we must be able to explain the origin of the internal velocity
dispersion and how it depends on the size of the region
considered (Larson 1981; Heyer \& Brunt 2004). 
The evolution of the velocity dispersion as a function
of the size of the region considered is shown in Figure~3 (from Bonnell \etal 2006). 
The initially low velocity dispersion, of order the sound
speed $v_s \approx 0.6$ km/s, increases as the gas passes through the
spiral shock. At the same time, the velocity dispersion increases more
on larger scales, producing a $v_{\rm disp} \propto R^{0.5}$  velocity
dispersion size-scale relation. 
The basic idea is that when structure exists
in the pre-shocked gas, the stopping point of a particular clump depends
on the density of gas with which it interacts.  thus some regions will
penetrate further into the shock, broadening it and leaving it with a
remnant velocity dispersion in the shock direction. Smaller scale regions  
in the shock are likely to have more uniform
momentum injection as well as encountering similar amounts of mass. This 
then results in small velocity dispersions. Larger
regions will have less correlation in both the momentum injection and mass loading 
such that there will be a larger dispersion in the post-shock velocity. Any clumpy
shock can induce such velocity dispersions. Thus the fractal nature
of the ISM passing through a spiral shock is a straightforward explanation for how the velocity dispersion size
relation arises in molecular clouds (Dobbs \& Bonnell 2006b).

\subsection{The clump-mass spectrum}

Clumpy shocks may also be an important role in setting the clump-mass spectrum.
Numerical simulations of colliding clumpy flows show that from an initial
population of identical clumps, the shocked gas contains a spectrum of clump masses
that is consistent with a Salpeter-like slope (Clark \& Bonnell 2006). This clump mass spectrum arises due to the coagulation
and fragmentation of the clumps in the shock, and is very similar to that observed in dense
prestellar cores (Motte \etal 1998). The relation between this clump-mass spectrum and the stellar
IMF is unclear as it arises independently of self-gravity and thus  a one-to-one
mapping of clump to stellar masses is unlikley. instead, the produced clumps
are likely to be a combination of unbound clumps that will not form stars and clumps
that are very bound that will form multiple many stars.

\section{Conclusions}

The triggering of star formation by the passage of clumpy gas through
a spiral arm can explain  many of the observed properties of
star forming regions.  Molecular cloud formation occurs as long
as the pre-shock gas is cold ($T \simless 100$ K).  
The shock forms dense structures in the gas which become
locally bound and collapse to form stars. The clouds are globally
unbound and thus disperse on timescales of $10^7$ years, resulting in
relatively low star formation efficiencies. 
In addition, the clumpy shock reproduces the
observed kinematics of GMCs, the so-called 'Larson' relation. There is
no need for any internal driving of the quasi-turbulent random
motions.  The internal structure of GMCs can also be understood as being
a produce of a clumpy shock and even the observed clump-mass spectrum in 
pre-stellar cores is reproduced. Finally, the clouds are sheared upon
leaving the spiral arms, producing the spurs and feathering commonly
observed in spiral galaxies.

\begin{discussion}

\end{discussion}


\begin{thebibliography}{}

\bibitem[]{} Baade, W., 1963, Evolution of stars and Galaxies,  Harvard University 
press (Cambridge), p 63.

%\bibitem[\protect\citeauthoryear{Bash, Green, \& Peters}{1977}]{1977ApJ...217..464B} Bash F.~N., Green E., Peters W.~L., 1977, ApJ, 217, 464 

\bibitem[]{} Bate M. R., Bonnell I. A., Price N. M., 1995, \mnras, 277, 362.

%\bibitem[]{} Benz W., Bowers R. L., Cameron A. G. W., Press B., 1991, \apj, 348, 647.

\bibitem[]{} Blitz, L. \& Williams, J., 1999, {\sl The origin of stars and planetary systems}, eds C.J.Lada, N.D. Kylafis, (Kluwer:Dordrecht),  3

\bibitem[]{bonnell}
Bonnell, I.A., Dobbs, C.L., Robitaille, T.P., Pringle, J.E., 2006, MNRAS, {\bf 365}, 37

\bibitem[]{} Clark, P.C.,  Bonnell, I.A., 2004 \mnras, {\bf 347}, L36
 
\bibitem[]{} Clark, P.C.,  Bonnell, I.A., 2006 \mnras, {\bf 368},  1787

\bibitem[]{} Clark, P.C.,  Bonnell, I.A., Zinnecker, H., Bate, M.R., 2004 \mnras, {\bf 359}, 809

\bibitem[\protect\citeauthoryear{Cowie}{1981}]{1981ApJ...245...66C} Cowie 
L.~L., 1981, ApJ, 245, 66  

\bibitem[]{} Dobbs C.L., Bonnell I.A., 2006a, MNRAS, {\bf 367}, 873

\bibitem[]{} Dobbs C.L., Bonnell I.A., 2006b, MNRAS, in press

\bibitem[]{} Dobbs C.L., Bonnell I.A., Pringle J.E., 2006, MNRAS, {\bf 371}, 1663

\bibitem[\protect\citeauthoryear{Elmegreen}{1991}]{1991ApJ...378..139E} 
Elmegreen B.~G., 1991, ApJ, 378, 139 
 
\bibitem[\protect\citeauthoryear{Elmegreen \& 
Elmegreen}{1983}]{1983MNRAS.203...31E} Elmegreen B.~G., Elmegreen D.~M., 
1983, MNRAS, 203, 31 

\bibitem[]{} Elmegreen B., Scalo, J., 2004, ARA\&A, {\bf 42},  211

\bibitem[\protect\citeauthoryear{Ferguson et 
al.}{1998}]{1998ApJ...506L..19F} Ferguson A.~M.~N., Wyse R.~F.~G., 
Gallagher J.~S., Hunter D.~A., 1998, ApJ, 506, L19 

\bibitem[]{}Heyer M. H., Brunt C. M., 2004, ApJ, 615, 45

\bibitem[\protect\citeauthoryear{Ferguson et 
al.}{1998}]{1998ApJ...506L..19F} Ferguson A.~M.~N., Wyse R.~F.~G., 
Gallagher J.~S., Hunter D.~A., 1998, ApJ, 506, L19 


\bibitem[]{} Larson R. B., 1981, MNRAS, 194, 809.

\bibitem[]{} Mac Low, M.M., Klessen, R.S., 2004, RvMP, {\bf 74}, 125

\bibitem[]{} Monaghan J. J., 1992, ARA\&A, 30, 543.

\bibitem[]{} Motte F., Andr\'e P., Neri R., 1998, A\&A, {\bf 336}, 150

\bibitem[]{} Roberts, W.W., 1969,  \apj, {\bf 158}, 123
 



\end{thebibliography}
\end{document}